\documentclass[12pt,preprint]{aastex}

\def\msun{{\,M_\odot}}

\slugcomment{Accepted by ApJ Lett}

\shorttitle{Djorgovski et al.}
\shortauthors{Collapsed Cores in Globular Clusters}

\begin{document}


\title{A Radio Outburst Nearly Coincident with the Large X-ray Flare from Sgr A* on 
2002-10-03}

\author{Jun-Hui Zhao\altaffilmark{1}, R. M. Herrnstein\altaffilmark{1,2},
G. C. Bower\altaffilmark{3}, W. M. Goss\altaffilmark{4} and S. M. 
Liu\altaffilmark{5}}

\altaffiltext{1}{Harvard-Smithsonian CfA, 60 Garden St., MS 78, Cambridge, MA02138.
jzhao@cfa.harvard.edu}
\altaffiltext{2}{
Dept. of Astronomy, Columbia University, Mail Code 5246, 550 West 120th St. New York, NY
}
\altaffiltext{3}{601 Campbell Hall, Radio Astronomy Lab, UC Berkeley
    Berkeley, CA 94720}
\altaffiltext{4}{NRAO, P. O. Box 0, Socorro, NM87801}
\altaffiltext{5}{Center for Space Science and Astrophysics, Stanford University, Stanford, CA 
94305-4060}

\begin{abstract}

A large radio outburst from Sgr A* was observed during the VLA weekly monitoring program at 2
cm, 1.3 cm and 7 mm, nearly coincident with the brightest X-ray flare detected to date
with the XMM-Newton X-ray Observatory on 2002-10-03. The flux density of 1.9$\pm$0.2
Jy measured at 7 mm exceeds the mean value (1.00$\pm$0.01 Jy) by a factor 
of $\sim 2$,  one of the two
highest increases observed during the past three years (June 2000-October 2003), 
while less significant increases in
flux densities were observed at 1.3 cm and 2 cm. The radio observation started 13.5 hrs
after the onset of the X-ray flare (which had occurred over a 45 min
duration) and continued for 1.3 
hrs. During the observation, there was no significant ($<3\sigma$)  change in the
radio flux densities at all the three 
wavelengths, indicating that the radio outburst varied on a
timescale of $>1$hr. A spectral index of 
$\alpha=2.4^{+0.3}_{-0.6}$ (${\rm S \propto\nu^\alpha}$) was derived for the outburst component, 
consistent with an optically thick nonthermal synchrotron source. 
These results suggest that energetic 
electrons responsible for the radio outburst might be produced via a process associated 
with the X-ray flare, then transported to large radii, producing the observed radio 
outburst. The observation is the first evidence for a correlated variation in  
the radio and X-ray emissions from Sgr A*.

\end{abstract}


\section{Introduction}
Radio  ``flares'' or outbursts (hereafter) on a timescale of a few days to weeks 
have been observed at short wavelengths from a few centimeters  to 1 mm
from Sgr A*, the compact radio source in the Galactic Center 
\citep{zha1992, wri1993, miy1999, zha2001, zha2003, her2003}. 
Radio observations have shown variations as short as a few hours
\citep{bow2002}.
With 
the {\it Chandra} and {\it XMM-Newton} observatories, X-ray flares with shorter variation time 
($\sim 1$ hr) have been frequently observed from the direction of Sgr A* over the past two 
years \citep{bag2001, bag2003, gol2003, por2003}. Because both 
radio and X-ray emissions are 
presumably produced close to the black hole, where energetic electrons responsible 
for the emissions are produced due to dissipation of the gravitational energy of the 
accreting plasma, the lack of correlated variation between radio and X-ray emission
 poses a number of 
theoretical challenges \citep{bag2003}. Recently, SgrA* has been detected in the IR band
showing the variation in  flux density on two 
timescales: (1)  a few days to a week
similar to that of the radio outbursts \citep{ghe2003b} and (2) $\sim$1hr for the
IR flares (similar to X-ray flares) \citep{gen2003}. 

On 2003-10-03, a giant X-ray flare from Sgr A* was detected with a peak luminosity 160
times higher than the quiescent value \citep{por2003}. No X-ray flare with such a high
luminosity has been previously detected. With the high sensitivity of {\it XMM-Newton}, it was
shown that the flare was variable on a time scale of 200 seconds, corresponding to an 
emission region no larger than ${\rm 5 r_S}$, where ${\rm r_S=1.2\times10^{12}}$ cm is the Schwarzschild 
radius of the black hole with a mass
of  $4\times10^6\msun$ \citep{ghe2003a}. Moreover, the flare has a soft 
photon spectrum with an spectral index of $\Gamma=2.5\pm0.3$, distinguishing itself from 
other relatively weaker  and 
more frequent X-ray flares with $\Gamma = 1.3^{+0.5}_{-0.6}$ \citep{bag2003}.

Coincidentally, one observation of the weekly VLA monitoring program was carried out 
$\sim$0.5 day 
after the X-ray observation and a radio outburst was detected at 7 mm. 
In this letter, we report detailed properties of the radio outburst on 2002-10-03. The radio
observations in the epochs close to the other weaker X-ray flares are also presented.

\section{Observations \& Calibrations}
The observations and data reduction for the VLA weekly monitoring program are discussed and 
reported by Herrnstein et al. (2003). On 2002-10-03, Sgr A* was observed at 2 cm, 1.3 cm and 
7 mm with a total BW of 100 MHz. The total observation time was 2 hrs 
including observations of a primary calibrator 3C 286, and QSOs 1741-312, 1817-254, 1730-130 
for phase and amplitude corrections. Scans on calibrator 1817-254 were interleaved between 
Sgr A*'s scans ($\sim 5$ mins each) at the three observing wavelengths. Three scans 
were made on both Sgr A* and 1817-254 at each wavelength over a period of 1.3 hrs. 

The absolute flux scale was determined from 3C 286. 
The final fractional uncertainties from a conservative estimate 
for the day-to-day variability 
are 6.1\%, 6.2\% and 9.2\% at 2 cm, 
1.3 cm and 7 mm, respectively \citep{her2003}.
The intra-hour variability of 1817-254 appeared to be small 
during the observation on  2002-10-03. Any variation was $<$ 5\%  
during the 1.3 hr at 7 mm and $<$ 2\% at 2 and 1.3 cm.
In addition to measurements on 2002-10-03, measurements of radio flux densities
at the three wavelengths corresponding to other X-ray events with a change  
in amplitude by a factor of 20 or greater were also made.

\section{Results}


Figure 1 shows measurements of flux densities  at 2 cm, 1.3 cm 
and 7 mm  
obtained from the weekly monitoring program \citep{her2003}
one month before and after the 2002-10-03 X-ray event. 
The flux density at 7 mm showed at
least a 4$\sigma$ increase on 2002-10-03. 
Variations in flux density at 1.3 cm and 2 cm were 3$\sigma$ and $<1\sigma$,
respectively.

The sampling period of 
the VLA monitoring program was about 1 
week. The measurements a week before and after 
the outburst showed that the flux densities were 
consistent with the  mean value of the flux densities averaged 
over the past three years (June 2000  - October 2003; see Table 1).  
Therefore, the timescale of the radio outburst at 7 mm must be less than 
two weeks for this event.  The
X-ray light curve observed with XMM-Newton showed a larger flare (160 x) \citep{por2003}. 
The X-ray observation started at
UT 7$^h$ 18$^m$ 8$^s$  on 2002-10-3 (Porquet 2003,
private communication). The X-ray flare onset occurred at UT 10$^h$ 5$^m$ and
lasted for about 45 mins. Our VLA observation started  on 2002-10-03 at 23$^h$ 30$^m$, 
about 13.5 hrs after the onset of the X-ray flare, and ended on 2002-10-04 at 0$^h$ 45$^m$
with three 5min scans on Sgr A* and the calibrator. Based on
the three measurements averaged on each scan over a 1.3 hr, we 
found that there were 
no significant variations in flux density ($< 3\sigma$)
 during the VLA observation at all the three observing 
wavelengths. A time scale for significant 
change ($> 3\sigma$) in radio flux density 
was likely  greater than 1 hr for the radio outburst.

Table 1 summarizes the radio property for the radio outburst 
corresponding to the X-ray flare on 2002-10-03.
The mean radio flux densities of ${\rm <S>}$ (Column 2) and the total 
flux densities ${\rm S_{t}}$ observed on 2002-10-03 
(Column 3) for the three observing wavelengths
are given in Column 2 and 3 respectively.
The flux densities for the outburst component ( ${\rm \Delta S
= S_{t} -<S>}$) are given in column 4.
The spectral index of $\alpha =0.71\pm$0.11 (${\rm S \propto \nu^\alpha}$)
derived from the total flux densities
appears to be significantly greater than the spectral index of  
0.2$\pm$0.02
derived from the mean flux densities  
observed over the past three years (June 2000 - October 2003).
The increase in the spectral index of the 
2002-10-03 event is consistent with the general 
correlation between spectral index and flux 
density at 7 mm \citep{her2003}.  A spectral 
index of $\alpha \approx 2.4^{+0.3}_{-0.6}$ was 
derived for the outburst component, which is consistent with
an optically thick, nonthermal synchrotron source.

In addition to the 2002-10-03 X-ray event
Fig. 2 shows a fraction of radio light curves covering a period $\pm$30 d for each of
the other three X-ray flares (a factor of 20 greater than a quiescent level)
detected in the past three years. 

For the 2000-10-27 X-ray flare observed 
with Chandra \citep{bag2001}, the peak X-ray flux was a factor of 45 greater than the quiescent
level. Radio observations
made three days before and four days after the X-ray event showed
no significant increase in radio flux density  measured 
in the two epochs close to this large X-ray event.
However,  the flux densities measured 9 days after the X-ray event showed 
$\sim3\sigma$ increase
in the radio flux densities at 2 and 1.3 cm and $\sim2\sigma$ at 7mm.

For the 2001-09-04 event, the beginning phase of this X-ray flare was
observed with XMM-Newton \citep{gol2003};  the amplitude of this flare 
was a factor of 20 greater than the quiescent
level. There were two observations at radio wavelengths carried
out  three and four days before the X-ray event. The flux densities  
were $2-3\sigma$s higher than
the mean flux density at all the  three observing wavelengths. The flux densities
measured five days after the X-ray event appeared
 to be consistent with the mean flux densities.

During the May 2002, there was a multi-wavelength campaign that coincided
with 100 hrs of Chandra observations. 
 During that period, the Chandra observations showed that flares with a factor
of 5 greater   in amplitude occurs once every day and larger flares with a factor
of 10 greater occurs once every two days \citep{bag2003}. 
A large flare (20x) occurred on 2002-05-29.
However, the radio measurements made
3 days before and 4 days after the 2002-05-29 event showed
 no significant variations
in flux density. There were also no significant variations observed at 7, 3 and 1.3 mm using VLBA, ATNF, and SMA,
respectively, in response to these multiple
small X-ray flares \citep{bag2003}.

The lack of correlation in radio flux density
at wavelengths ranging from 2 cm  to 7 mm with the weaker X-ray flares
has been puzzling. The weekly  VLA monitoring
program appeared to be not sensitive to  the weaker X-ray flares.
However, the large X-ray flare on 2002-10-03 appeared to be special in many aspects.
In addition to a high luminosity and softer X-ray spectrum, 
a large radio outburst at 
7mm was detected within the same day.
Based on the fact that we only detected two large  (2x) radio
outbursts at 7mm from observations of 121 epochs weekly sampled
over the past 3yr period between June 2000 to October 2003,
the probability of detecting such a large (2x)
radio outburst within a time interval of ${\rm \Delta t}$
appears to be ${\rm P_R \sim 0.01 { \Delta t \over \Delta t_{sampling}}}$,
where ${\rm \Delta t_{sampling}\sim 1 wk}$, the radio sampling interval. 
On the other hand, during the past three years overlapping with the period
of VLA  monitoring program, observations of a few hundred hours were carried out 
at X-ray with both Chandra and XMM-Newton to search for X-ray flares, 
but to date only one large (160x) X-ray flare has been detected. Considering
the typical duration of ${\rm\Delta t_X \sim 1}$hr for the  X-flares, then 
the detection probability of a large flare (160x) within 
${\rm \Delta t}$ appears 
to be P$_{\rm  x}< 0.01 {\rm \Delta t\over\Delta t_X}$.
A large uncertainty in estimate
of ${\rm P_X}$ is owing to the sparse observations at X-ray and a large chance
to have a stellar X-ray transient in the large XMM-Newton beam.
However, if both the radio and X-ray events were randomly produced from
two independent processes, then the probability of detecting large radio 
and X-ray events within a time-scale of the radio outburst
${\rm \Delta t_R}$
would be  
${\rm P = P_R\times P_X  < 0.0001{\Delta t_R^2 \over 
\Delta t_X \Delta t_{sampling}}}.$ Based on our observations, the
time-scale on the radio outburst is well constrained in a range of  
${\rm 1 hr < \Delta t_R \le  1wk}$. For
${\rm \Delta t_R \le 1wk}$, the probability 
is $<$2 \%. For  ${\rm \Delta t_R\sim}$a few days,
a reasonable guess for the time-scale of the radio outburst,
then the probability would be 0.1 - 0.3\%. 
Statistically, we  have a good confidence to reject the hypothesis
that the two events were produced from two independent random processes. 
The two large events observed at radio and X-ray appeared to be related.

Since the radio outburst component of the 2002-10-03 event had a spectral
index of $2.4^{+0.3}_{-0.6}$, it is likely that the large X-ray flare
was associated with  an optically thick, nonthermal synchrotron  component
in Sgr A*. 
With an upper limit of 8 km s$^{-1}$ for the intrinsic proper motion
of Sgr A* determined from VLBI measurements, a lower limit on the mass 
of Sgr A* of 4$\times10^5$ M$_\odot$ \citep{rei2003} is placed.
This limit along with the compactness of Sgr A* with an intrinsic
size of 0.24 mas or 24  ${\rm r_S}$\citep{bow2003}
suggests that Sgr A* is very likely associated with the putative 
supermassive black hole at the Galactic center.
If the radio outburst component was confined within ${\rm < 24 r_S}$ 
from the black hole, 
then the X-ray flares must arise 
from the inner region of the accretion flow near the event horizon of
the SMBH rather than from star-star collision \citep{bag2001} or from  the heated part of
the disk via star-disk interaction \citep{nay2003}.

\section {Discussion}

A radio outburst observed at 7 mm coinciding in about 0.5 day 
with  the most luminous X-ray flare 
appeared to show an intimate relation between the two events.
The significant variability on a timescale of 200s observed during the X-ray flare
indicates that the size of the X-ray emitting region  is about ${\rm 5 r_S}$.
The observed optically thick nonthermal outburst component was likely produced from a 
region with a size ${\rm >5 r_S}$ as suggested by the following facts.

The lack of significant  variation
in flux density during the radio observation of the outburst 
suggests that its variation time scale should be longer than one hour, which is about twenty 
times greater than the variation timescale  of the X-ray flare.

If the outburst component is indeed a self-absorbed nonthermal synchrotron
source with a turnover frequency $\nu_m>43$ GHz and a peak
flux density ${\rm S_m\>\sim0.86 Jy(\nu_m/43\ {\rm 
GHz})^{2.5}}$. The turnover frequency
is likely ${\rm \nu_m\sim300}$GHz 
(or $\lambda_m\sim1$mm) based on previous observations of Sgr A* at submillimeter
wavelengths \citep{zha2003, ser1997}. 
The brightness temperature of the outburst component can be calculated:
\begin{equation}{\rm 
T_B \approx 3.2\times 10^{11} {\rm K}\left({\rm \nu_m
\over {\rm 43 GHz}}\right)^{0.5}\left({D\over 8\ {\rm kpc}}\right)^2
\left({5 r_S\over 
d}\right)^2,}
\end{equation}
where D is the distance to Sgr A* and d is the diameter of the source. 
The brightness 
temperature could break the inverse Compton scattering
 limit for 
self-absorbed synchrotron source 
\citep{rea1994, sin1994} if $\nu_m\sim$300 GHz and  d$\sim$5 ${\rm r_S}$. 
If the onset of 
the radio outburst indeed arose from the X-ray region with a
size of d${\rm \sim5~r_S}$, 
the outburst component must expand substantially to reach a size 
of d${\rm \sim24~r_S}$  as observed with VLBI\citep{bow2003} so that
 a drastic energy loss 
from the self-synchrotron inverse Compton scattering (SSC) 
lasted only for a short period,
perhaps, of $\sim$1 hr as suggested by the duration 
of the X-ray flare.

On the other hand, for a spherical,  
optically thick synchrotron source,
the magnetic field (B) 
can be estimated from the source angular size 
(${\rm \theta={d\over D}}$), ${\rm S_m}$
and ${\nu_m}$ \citep{mar1983}: 
\begin{equation}{\rm
B\approx 0.035 G\left({8\ {\rm kpc}\over D}\right)^4\left({d\over 5 r_S}\right)^4
\left({\rm \nu_m\over 43GHz}\right)^5 \left({\rm S_m\over 1Jy}\right)^{-2}
\approx 0.045 G \left({8\ {\rm kpc}\over D}\right)^4\left({d\over 5 r_S}\right)^4}
\,.
\end{equation}
For d${\rm \sim24~r_S}$, the typical size of Sgr A* as measured with
VLBI at 7mm, the inferred B$\sim24$ G
is consistent 
with the characteristic magnetic field near 
the black hole as suggested in theory \citep{liu2002b}. 
For B$\sim24$ G,
the  synchrotron cooling time at 43 GHz of $\sim$0.5 d is inferred,
which is consistent with our observations of no significant
variations in flux density at 7mm within the observing interval of 1 hr.

Thus, the 7 mm radio outburst was likely produced at a relatively 
large size scale 
with respect to the source size of the X-ray flare. Given that the radio 
outburst was observed about 13.5 hrs 
after the onset of the X-ray flare, it is reasonable to suggest that 
the electrons 
producing the radio outburst were energized \citep{yua2003}
via a process 
associated with the X-ray flare
 and transported to larger radii via a diffusion process \citep{liu2002b, zha2003}
if not a collimated jet \citep{fal1993, yua2002}. 
The correlation between radio spectral index and 7mm flux density observed with the
VLA \citep{her2003} and the frequent X-ray flares observed with Chandra 
\citep{bag2001, bag2003}
may also suggest that the nonthermal 
electrons responsible for the overall  
radio emission from Sgr A* are 
probably energized via a 
similar process.

To justify the above scenario for correlated variation between 
radio and X-ray emissions,
the energetic electrons diffusing outward to a large radii must  
contain enough energy
in order to sustain the radio outburst for a few days. Because the 
X-ray flare was 
probably produced 
via SSC in the model, to avoid drastic inverse Compton losses, 
the magnetic field energy 
density must be larger than the photon energy density near 
the black hole. If the nonthermal 
electrons were still in energy equipartition with the magnetic field, 
the energy flux 
associated with the escaping nonthermal electrons should be larger 
than the energy flux 
of the photons. The X-ray luminosity between $2-10$ keV during the 
flare was about $4\times 
10^{35}$ erg s$^{-1}$. Given the softness of the X-ray emission, the 
total X-ray luminosity 
could be around $10^{36}$ ergs s$^{-1}$. Thus, the 
total energy X-ray emission produced during 
the X-ray flare was about $10^{40}$ ergs. The energy carried  
by nonthermal electrons 
during the X-ray flare was then about $10^{40}$ ergs, 
which could sustain 
an outburst component of 0.86 Jy at 7 mm for a few days.

\section {Conclusions}

One of the two strongest radio outbursts observed over the 
past three years appears to
be related to the largest
X-ray flare to date which was observed on 2002-10-03. The  
correlation between
the strong emissions at X-ray and radio suggests that the radio outbursts are 
powered  via an electron acceleration process during the X-ray flare.
Our observations and analysis 
are consistent with the hypothesis that the X-ray flare 
originates from self-synchrotron inverse Compton scattering
process  close to the supermassive black hole at the 
Galactic center.

\acknowledgments
We thank Delphine Porquet and Fulvio Melia for their communication and 
 X-ray data.
The VLA is operated by the 
NRAO. The NRAO is a facility of the National Science Foundation operated 
under
cooperative
agreement by Associated Universities, Inc.

\clearpage
\figcaption[fig1.ps]{Top: The flux density measurements
made from the VLA observations 
at 2 cm (open squares), 1.3 cm (open diamonds) and 7 mm (solid dots) 30 days 
before-and-after the 2002-10-03 X-ray flare. The error bars mark
the 1$\sigma$ uncertainty in the variability measurements.
The horizontal lines (solid, dash, and dash-dot-dash) mark
the mean values at 7 mm, 1.3 and 2 cm, respectively.
\label{fig1}}

\figcaption[fig2.ps]{The flux density measurements
made from the VLA observations
at 2 (open squares), 1.3 cm (open diamonds) and 7 mm (solid dots) 30 days 
before-and-after the three X-ray flares detected on 2000-10-27 (Top),
2001-09-04 (Middle) and 2002-05-29 (Bottom).
The horizontal lines (solid, dash, and dash-dot-dash) mark
the mean values at 7 mm, 1.3 and 2 cm, respectively.
The X-ray flares were observed
with peak flux a factor of 20 greater than the value of
the quiescent state. The error bars mark
the 1$\sigma$ uncertainty in the variability measurements.
The origin of the horizontal axis corresponds to  the date of each x-ray flare.
\label{fig2}}

\newpage
 \plotone{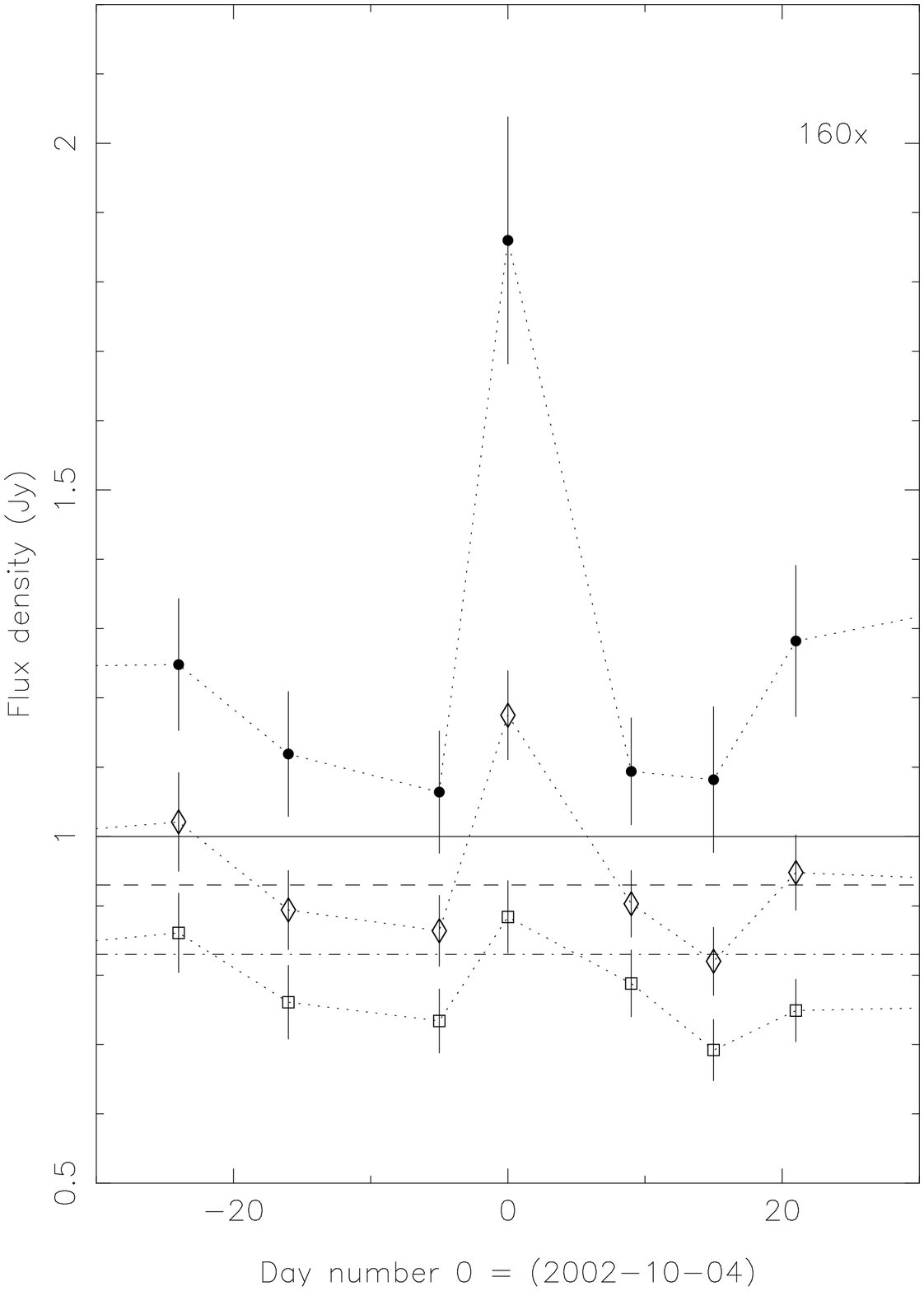}
 \newpage
 \plotone{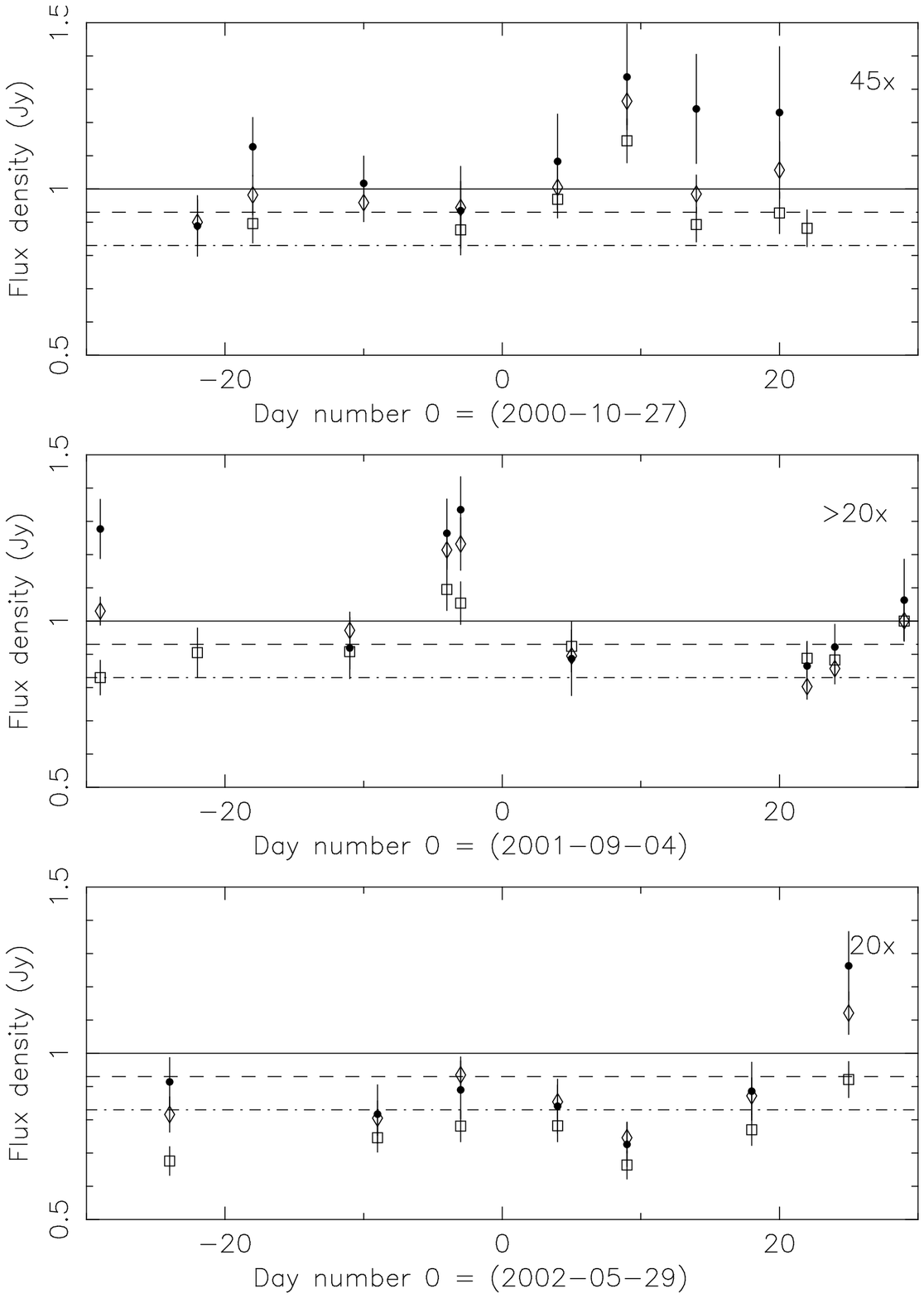}

\clearpage

\begin{deluxetable}{cccc}
\tablecaption{Radio Properties of The Outburst on 2002-10-03}
\tablewidth{0pt}
\tablehead{
\colhead{$\lambda$ (cm)}          & 
\colhead{${\rm <S>}$ (Jy)}        & 
\colhead{S$_{\rm t}$ (Jy)}   &
\colhead{${\rm \Delta S}$ (Jy)}
}
\startdata
2.0 & 0.83$\pm$0.01 & 0.88$\pm$0.05  & 0.05$\pm$0.05 \\
1.3 & 0.93$\pm$0.01 & 1.18$\pm$0.06  & 0.25$\pm$0.06 \\
0.7 & 1.00$\pm$0.01 & 1.86$\pm$0.18  & 0.86$\pm$0.18 \\
\cutinhead{$\alpha$ (S$\propto\nu^\alpha$)}
2.0/1.3/0.7& 0.20$\pm$0.02 & 0.7$\pm$0.1 & 2.4$^{+0.3}_{-0.6}$ \\
\enddata

\end{deluxetable}


\end{document}